\documentclass[12pt]{article}
\usepackage{fancyhdr}

\usepackage{mathrsfs}
\usepackage[T1]{fontenc}
\usepackage{mathpazo}
\usepackage{setspace}
\usepackage{amsfonts}
\usepackage{amssymb}
\usepackage{amsmath}
\usepackage{epsfig}
\usepackage{latexsym}
\usepackage{color}
\usepackage{graphicx}
\usepackage{nicefrac}
\usepackage[latin1]{inputenc}
\usepackage{slashed}
\usepackage{multirow}
\usepackage{comment}
\usepackage{soul}
\usepackage{hyperref}
\usepackage{cite}

\usepackage[titletoc]{appendix}
\def\hybrid{\topmargin -20pt    \oddsidemargin 0pt
        \headheight 0pt \headsep 0pt
        \textwidth 6.25in       % A4 paper
        \textheight 9.25in       % A4 paper
        \marginparwidth .875in
        \parskip 5pt plus 1pt   \jot = 1.5ex}

\hybrid

\def\baselinestretch{1.2}

\catcode`\@=11

\def\marginnote#1{}
\newcount\hour
\newcount\minute
\newtoks\amorpm
\hour=\time\divide\hour by60
\minute=\time{\multiply\hour by60 \global\advance\minute by-\hour}
\edef\standardtime{{\ifnum\hour<12 \global\amorpm={am}%
        \else\global\amorpm={pm}\advance\hour by-12 \fi
        \ifnum\hour=0 \hour=12 \fi
        \number\hour:\ifnum\minute<10 0\fi\number\minute\the\amorpm}}
\edef\militarytime{\number\hour:\ifnum\minute<10 0\fi\number\minute}
\def\draftlabel#1{{\@bsphack\if@filesw {\let\thepage\relax
   \xdef\@gtempa{\write\@auxout{\string
      \newlabel{#1}{{\@currentlabel}{\thepage}}}}}\@gtempa
   \if@nobreak \ifvmode\nobreak\fi\fi\fi\@esphack}
        \gdef\@eqnlabel{#1}}
\def\@eqnlabel{}
\def\@vacuum{}
\def\draftmarginnote#1{\marginpar{\raggedright\scriptsize\tt#1}}

\def\draft{\oddsidemargin -.5truein
        \def\@oddfoot{\sl preliminary draft \hfil
        \rm\thepage\hfil\sl\today\quad\militarytime}
        \let\@evenfoot\@oddfoot \overfullrule 3pt
        \let\label=\draftlabel
        \let\marginnote=\draftmarginnote
   \def\@eqnnum{(\theequation)\rlap{\kern\marginparsep\tt\@eqnlabel}%
\global\let\@eqnlabel\@vacuum}  }

\def\preprint{\twocolumn\sloppy\flushbottom\parindent 2em
        \leftmargini 2em\leftmarginv .5em\leftmarginvi .5em
        \oddsidemargin -.5in    \evensidemargin -.5in
        \columnsep .4in \footheight 0pt
        \textwidth 10.in        \topmargin  -.4in
        \headheight 12pt \topskip .4in
        \textheight 6.9in \footskip 0pt
        \def\@oddhead{\thepage\hfil\addtocounter{page}{1}\thepage}
        \let\@evenhead\@oddhead \def\@oddfoot{} \def\@evenfoot{} }

\def\numberbysection{\@addtoreset{equation}{section}
        \def\theequation{\thesection.\arabic{equation}}}

\def\underline#1{\relax\ifmmode\@@underline#1\else
        $\@@underline{\hbox{#1}}$\relax\fi}

\def\titlepage{\@restonecolfalse\if@twocolumn\@restonecoltrue\onecolumn
     \else \newpage \fi \thispagestyle{empty}\c@page\z@
        \def\thefootnote{\fnsymbol{footnote}} }

\def\endtitlepage{\if@restonecol\twocolumn \else \newpage \fi
        \def\thefootnote{\arabic{footnote}}
        \setcounter{footnote}{0}}  %\c@footnote\z@ }

\catcode`@=12
\relax

\def\figcap{\section*{Figure Captions\markboth
        {FIGURECAPTIONS}{FIGURECAPTIONS}}\list
        {Figure \arabic{enumi}:\hfill}{\settowidth\labelwidth{Figure
999:}
        \leftmargin\labelwidth
        \advance\leftmargin\labelsep\usecounter{enumi}}}
 \relax
\def\tablecap{\section*{Table Captions\markboth
        {TABLECAPTIONS}{TABLECAPTIONS}}\list
        {Table \arabic{enumi}:\hfill}{\settowidth\labelwidth{Table
999:}
        \leftmargin\labelwidth
        \advance\leftmargin\labelsep\usecounter{enumi}}}
 \relax
\def\reflist{\section*{References\markboth
        {REFLIST}{REFLIST}}\list
        {[\arabic{enumi}]\hfill}{\settowidth\labelwidth{[999]}
        \leftmargin\labelwidth
        \advance\leftmargin\labelsep\usecounter{enumi}}}
 \relax
\makeatletter
\newcounter{pubctr}
\def\publist{\@ifnextchar[{\@publist}{\@@publist}}
\def\@publist[#1]{\list
        {[\arabic{pubctr}]\hfill}{\settowidth\labelwidth{[999]}
        \leftmargin\labelwidth
        \advance\leftmargin\labelsep
        \@nmbrlisttrue\def\@listctr{pubctr}
        \setcounter{pubctr}{#1}\addtocounter{pubctr}{-1}}}
\def\@@publist{\list
        {[\arabic{pubctr}]\hfill}{\settowidth\labelwidth{[999]}
        \leftmargin\labelwidth
        \advance\leftmargin\labelsep
        \@nmbrlisttrue\def\@listctr{pubctr}}}
 \relax
\makeatother
\newskip\humongous \humongous=0pt plus 1000pt minus 1000pt

\newif\ifdtup

\relax

\def\be{\begin{equation}}
\def\ee{\end{equation}}
\def\ba{\begin{eqnarray}}
\def\ea{\end{eqnarray}}

\def\del{\partial}

\def\a{\alpha}

\def\b{\beta}

\def\d{\delta}
\def\D{\Delta}

\def\p{\pi}

\def\th{\theta}

\def\m{\mu}
\def\n{\nu}
\def\om{\omega}

\def\l{\lambda}
\def\L{\Lambda}
\def\s{\sigma}

\def\cL{{\cal L}}
\def\cR{{\cal R}}

\def\no{\noindent}

\def\qq{\qquad}

\def\IR{\relax{\rm I\kern-.18em R}}

\def \ha {{1\over 2}}

\def \ov {\over}

\def\diag{{\rm diag}}

\def\IR{\relax{\rm I\kern-.18em R}}
\def\IL{\relax{\rm I\kern-.18em L}}

\def\inv{^{\raise.15ex\hbox{${\scriptscriptstyle -}$}\kern-.05em 1}}

\def\cL{{\cal L}}
\def\cR{{\cal R}}

\def\Tr{{\rm Tr}}

\begin{document}
%Text\fontsize{13}{12}\selectfont Text

%\renewcommand{\theequation}{\arabic{equation}}
\renewcommand{\theequation}{\thesection.\arabic{equation}}
\csname @addtoreset\endcsname{equation}{section}

\newcommand{\beq}{\begin{equation}}
\newcommand{\eeq}[1]{\label{#1}\end{equation}}
\newcommand{\ber}{\begin{equation}}
\newcommand{\eer}[1]{\label{#1}\end{equation}}
\newcommand{\eqn}[1]{(\ref{#1})}
\begin{titlepage}
\begin{center}

%\hfill CALT-xx-yyyy\\
%\vskip -.1 cm
%\hfill hep--th/yymmnnn\\

${}$
\vskip .2 in

{\large\bf
%A new class of integrable deformations of current algebra theories}
A new class of integrable deformations of CFTs}

\vskip 0.4in

{\bf George Georgiou$^1$\ and \ Konstantinos Sfetsos}$^{2}$
\vskip 0.1in

\vskip 0.1in
{\em
${}^1$Institute of Nuclear and Particle Physics,
\\ National Center for Scientific Research Demokritos,
\\
Ag. Paraskevi, GR-15310 Athens, Greece
}
\vskip 0.1in

 {\em
${}^2$Department of Nuclear and Particle Physics,\\
Faculty of Physics, National and Kapodistrian University of Athens,\\
Athens 15784, Greece\\
}

\vskip 0.1in

{\footnotesize \texttt georgiou@inp.demokritos.gr, ksfetsos@phys.uoa.gr}

%\today

\vskip .5in
\end{center}

\centerline{\bf Abstract}

\no
We construct a new class of integrable $\s$-models based on current algebra
theories for a general semisimple group $G$ by utilizing a left-right asymmetric gauging.
Their action can be
thought of as the all-loop effective action of two independent WZW models
for $G$ both at level $k$,
perturbed by current bilinears mixing the different WZW models. A non-perturbative symmetry in the couplings parametric space is revealed. We perform the Hamiltonian analysis of the action and demonstrate integrability in
several cases.
We extend our construction to deformations of $G/H$ CFTs
and show integrability when $G/H$ is a symmetric space.
Our method resembles that used for constructing the $\l$-deformed integrable $\s$-models, but the results are distinct and novel.

\vskip .4in
\noindent
\end{titlepage}
\vfill
\eject

\newpage
%\vskip .3in

\tableofcontents

\noindent

\def\baselinestretch{1.2}
\baselineskip 20 pt
\noindent

%%%%%%%%%%%%%%%

\setcounter{equation}{0}
\section{Introduction }

We are interested in deformations of two-dimensional conformal field theories (CFTs)
which preserve integrability for all values of the deformations parameters.
This can serve as a basis for constructing generalizations of the original AdS/CFT correspondence simultaneously maintaining one of its key features, i.e. integrability.
In particular, in this work we will construct and study a new class of integrable two-dimensional $\s$-models based on current algebra theories for a general semisimple group $G$ having a Lagrangian realization in terms of  two independent WZW models both at level $k$.

\no
The strategy for constructing these new deformed integrable $\s$-models will
resemble that used for the $\l$-deformations \cite{Sfetsos:2013wia}, which
we briefly review next. This will also be useful in realizing
the similarities and differences of the two classes of models.
The starting point for constructing the $\s$-model for $\l$-deformations was to consider
the sum of the actions for the WZW model and the principal chiral model (PCM) for a group $G$ and then gauge
it. This action is given by
\ba
&& S_k(g,\tilde g, A_\pm) = S_{k}(g)
 +{k\ov \pi} \int d^2\s \ \Tr \big(A_- \del_+ g g^{-1} - A_+ g^{-1} \del_- g
+ A_- g A_+ g^{-1}  -  A_- A_+ \big)
\nonumber\\
&&\qq \qq \qq\qq -{1\ov \pi} \int d^2\s\
\Tr \big(t^a\tilde g^{-1} D_+ \tilde g)E_{ab}\Tr(t^b \tilde g^{-1} D_- \tilde g\big)\ .
\label{asygwzw}
\ea
The WZW action $S_{k}(g)$ for a group element $g\in G$  is given
by\footnote{
Our conventions are
\be
\s^\pm =\tau \pm \s  \ , \qq d\s^+ \wedge d\s^- = -2  d^2 \s \ ,\qq
d^2\s = d\tau \wedge d\s \ .
%\label{convwzw}
\ee}
\ba
S_{k}(g) = {k\ov 2\pi} \int d^2\s \Tr(\del_+ g^{-1} \del_- g)
+ {k\ov 12\pi} \int  \Tr(g^{-1} dg)^3\ .
\ea
The relative coefficient of the cubic term is completely dictated
by the Polyakov--Wiegman formula
\be
S_k(g_1 g_2) = S_k(g_1) + S_k(g_2)
- {k\ov \pi} \int d^2\s \ \Tr(g_1^{-1} \del_- g_1 \del_+ g_2 g_2^{-1})\ .
\label{polywig}
\ee
The first line of \eqn{asygwzw} is the gauged WZW for $G/G$, where the gauge
fields $A_\pm$ take values in the Lie algebra of $G$.  The second line in
\eqn{asygwzw} is the gauged PCM action for a group element $\tilde g\in G$ and
general coupling matrx $E_{ab}$.
The $t^a$'s are representation matrices obeying the Lie algebra
$[t_a,t_b]=i f_{abc} t_c$.
The coupling to the gauge fields is done, unlike the gauged WZW action, with minimal
coupling, i.e. $D_\pm \tilde g = \del_\pm\tilde g - A_\pm \tilde g$.
Both lines are separately invariant and so is the total action, under the transformation
\be
\d g = [g,u]\ ,\qq \d \tilde g = - u g \ ,\qq \d A_\pm = - \del_\pm u +[A_\pm ,u] \ ,
\label{traaasd}
\ee
where the infinitesimal parameter $u$ takes values in the Lie-algebra of $G$. Upon fixing
the gauge freedom as $\tilde g= \mathbb{I}$ and
integrating out the gauge fields one obtains the $\l$-deformed $\s$-model action
\be
S_{k,\l}(g)=  S_{k}(g)+ {k\ov \pi} \int d^2\s\ J^a_+ (\l^{-1}- D^T)^{-1}_{ab}J^b_-\ ,
\label{laact}
\ee
where the matrix
\be
\l= k ( k\mathbb{I} + E)^{-1}
\label{defl}
\ee
and
\be
\label{hg3}
J^a_+ = - i \Tr(t^a \del_+ g g^{-1}) ,\qq J^a_- = - i \Tr(t^a g^{-1}\del_- g )\ .
\qq D_{ab}= \Tr(t_a g t_b g^{-1})\ ,
\ee
The simplest case in which the matrix $\l$ is proportional to the identity is
integrable \cite{Sfetsos:2013wia}.
In addition, the action \eqn{laact}
exhibits a remarkable duality-type symmetry \cite{Itsios:2014lca,Sfetsos:2014jfa} given by
\be
k\to -k \ ,\qq \l \to  \l^{-1} \qq g\to g^{-1}  \ .
\label{duallli}
\ee
This symmetry was already argued to hold, using path integral arguments and manipulations in \cite{Kutasov:1989aw}. The action \eqn{laact} for small deformation parameters becomes
\be
\label{expan2kk}
S_{k,\l}(g) = S_{k}(g) + {k\ov \pi} \int d^2\s\ \l_{ab} J^a_{+} J^b_{-} + \cdots \ .
\ee
This construction was extended to cosets \cite{Sfetsos:2013wia,Hollowood:2014rla,Hollowood:2014qma} and supergroups \cite{Hollowood:2014rla,Hollowood:2014qma,Schmidtt:2016tkx} which are of particular interest in string theory and in the context of the gauge/gravity
correspondence. Since \eqn{laact} is no-longer corresponding to a conformal background, it is expected that the matrix elements $\l_{ab}$
will run under the renormalization group (RG) flow. Indeed,
the computation of the corresponding RG flow equations using gravitational methods was performed in
\cite{Itsios:2014lca,Sfetsos:2014jfa} and is in agreement with results from field theoretical  methods \cite{Kutasov:1989dt,Gerganov:2000mt,Appadu:2015nfa}. Furthermore, it was shown that these models, named generically as $\l$-deformed, can be embedded in specific
cases of low dimensionality to supergravity \cite{Sfetsos:2014cea,Demulder:2015lva,Borsato:2016zcf,Chervonyi:2016ajp,Chervonyi:2016bfl},
while their relation to $\eta$-deformations  for group and coset spaces introduced in \cite{Klimcik:2002zj,Klimcik:2008eq,Klimcik:2014}
and \cite{Delduc:2013fga,Delduc:2013qra,Arutyunov:2013ega} respectively, via Poisson-Lie T-duality
\cite{KS95a} and appropriate analytic continuations, was uncovered in \cite{Vicedo:2015pna,Hoare:2015gda,Sfetsos:2015nya,Klimcik:2015gba,Klimcik:2016rov}.
More recently, the computation of the all-loop correlators for the isotropic case, i.e. exact in $\l_{ab}=\l \d_{ab}$, but for $k\gg 1$,
 of current and primary field operators was performed \cite{Georgiou:2016iom, Georgiou:2015nka}. In these computations a few terms in perturbation theory obtained by CFT methods and the symmetry \eqn{duallli} were enough to obtain the results.
 Other selected and related recent works can be found in \cite{selected}.

\no
The plan of the paper is as follows:
In section 2, we present the construction of our model. This is achieved by utilizing a left-right asymmetric gauging of two independent WZW models both at level $k$ combined with two independent deformed PCMs. These gives rise to models characterized by the overall integer
coefficient $k$ and by two matrices $\l_1$ and $\l_2$.
In section 3, we prove that the theory is integrable for specific nono- or multi-parameter choices for the
deformation matrices $\l_1$ and $\l_2$. It turns out that, one constructs two independent Lax pairs, but
in order to show independence of the infinite conserved charges obtained from the corresponding monodromy matrices it is necessary
to perform a Hamiltonian analysis of the action and deal with the constrains that arise following the usual Dirac procedure.
A slight modification provides integrability for deformations corresponding to coset $G/H$ exact CFT provided $G/H$ is a symmetric space.
In section 4, we provide an explicit example corresponding to a deformation of an Abelian T-dual of the
$\s$-model corresponding to the direct product $SU(2)_k/U(1)\times SU(2)_k/U(1)$
exact CFT.
Finally, in section 5 we draw our conclusions and discuss some directions for future work.

\renewcommand{\theequation}{\thesection.\arabic{equation}}

\section{Constructing the new models}

The purpose of this section is to construct a new class of integrable models following in
spirit the construction presented in the previous section.
Consider replacing the usual gauged WZW action with \cite{Witten:1991mm}
\be
\begin{split}
&
S_k(g,A_\pm,B_\pm) = S_{k}(g)
 +{k\ov \pi} \int d^2\s \ \Tr \big(A_- \del_+ g g^{-1} - B_+ g^{-1} \del_- g
+ A_- g B_+ g^{-1}  
\\
&\qq\qq \qq\qq  -\ha  A_- A_+ - \ha B_+ B_- \big)\ ,
\end{split}
\ee
where now we use two different algebra valued fields $A_\pm $ and $B_\pm$.
The above action is not gauge invariant, but instead, under the infinitesimal transformation
\be
  \d g = g u_R - u_L g \ ,\qq
\d A_\pm =-\del_\pm u_L + [A_\pm, u_L]\ ,\qq
\d B_\pm =-\del_\pm u_R + [B_\pm, u_R]\ ,
\ee
with different infinitesimal parameters for the left and the
right transformations, transforms as
\be
\d S_k(g,A_\pm,B_\pm)  = {k\ov 2\pi} \int d^2\s \
\Tr\big[ (A_+ \del_- u_L - A_- \del_+ u_L) - (B_+ \del_- u_R - B_- \del_+ u_R)\big]\ .
\label{anogwzw}
\ee
This classical gauge anomaly is independent of the group element and this will be used to our advantage in our construction.
We are going to combine two such actions with two PCMs as follows
\be
\begin{split}
& S_{k}(g_1,g_2, \tilde g_1,\tilde g_2,A_\pm, B_\pm)= S_{k}(g_1,A_\pm,B_\pm)
+ S_{k}(g_2,B_\pm,A_\pm)
\label{fgje3}
\\
& \qq \qq\quad -{1 \ov \pi}   \int d^2\s \ \Tr \big(t^a \tilde  g_1^{-1} D_+ \tilde g_1)E_{1ab} \Tr(t^b \tilde g_1^{-1} D_- \tilde g_1\big)
\\
&\qq \qq\quad -{1 \ov \pi}   \int  d^2\s \
\Tr \big(t^a\tilde  g_2^{-1} D_+ \tilde g_2)E_{2ab} \Tr(t^b \tilde g_2^{-1} D_- \tilde g_2\big)\ ,
\end{split}
\ee
where we notice the interchange of the $A_\pm$ and $B_\pm$ in the two gauged WZW
actions. The covariant derivatives on $\tilde g_1$ and $\tilde g_2$ are defined as
$ D_\pm \tilde g_1 = \del_\pm \tilde g_1 - A_\pm \tilde g_1$ and
$ D_\pm \tilde g_2 = \del_\pm \tilde g_2 - B_\pm   \tilde g_2 $. The matrices $E_1$ and $E_2$ parametrizing the couplings.

\no
This action is invariant under the transformation
\ba
&&  \d g_1 =g_1 u_R - u_L g_1 \ ,\qq \d g_2 =  g_2 u_L - u_R g_2\ ,
\nonumber\\
&& \d \tilde g_1 =- u_L \tilde g_1 \ ,\qq \d g_2 =  - u_R \tilde g_2\ ,
\\
&&
\d A_\pm =-\del_\pm u_L + [A_\pm, u_L]\ ,\qq
\d B_\pm =-\del_\pm u_R + [B_\pm, u_R]\ .
\nonumber
\ea
Indeed, as in \eqn{asygwzw} the last two lines of \eqn{fgje3} concerning the PCMs are invariant on their own.
The using \eqn{fgje3} we have that
\ba
&& \d S_{k}(g_1,g_2, A_\pm, B_\pm) = {k\ov 2\pi}  \int d^2\s \
\Tr\big[ (A_+ \del_- u_L - A_- \del_+ u_L) - (B_+ \del_- u_R - B_- \del_+ u_R)\big]
\nonumber\\
&&
\phantom{xxxx}
+ {k\ov 2\pi}   \int d^2\s \
\Tr\big[ (B_+ \del_- u_R - B_- \del_+ u_R) - (A_+ \del_- u_L - A_- \del_+
 u_L)\big]= 0\ .
\ea
Then we may fix completely the gauge by choosing that
$\tilde g_1=\tilde g_2=\mathbb{I}$. In analogy to \eqn{defl} we introduce the definition
\be
\l_i  = k(k\mathbb{I} + E_i )^{-1} \ ,\qquad i=1,2\ ,
\ee
which actually are going to be the two deformation parameters.
Then, the gauge fixed action is given by
\be
\begin{split}
&  S_{k}(g_1,g_2, A_\pm, B_\pm) = S_{k}(g_1) + S_{k}(g_2)
 \\
 &\qq
 + {k\ov \pi} \int d^2\s \ \Tr \big(A_- \del_+ g_1 g_1^{-1}   - B_+ g_1^{-1} \del_- g_1
 + A_- g_1 B_+ g_1^{-1}  -   A_+ \l_1^{-1} A_-
\\
\label{gaufix}
 &\qq\qquad  +  B_- \del_+ g_2 g_2^{-1} -  A_+ g_2^{-1} \del_- g_2
 +  B_- g_2 A_+ g_2^{-1} -    B_+ \l_2^{-1} B_- \big)\ .
 \end{split}
\ee
Integrating out the gauge fields we find that
\be
\label{abp}
\begin{split}
&
A_+ = i (\mathbb{I}-\l_1^T D_1 \l_2^T D_2)^{-1}\l_1^T (J_{1+} +  D_1\l_2^T J_{2+})\ ,
\\
&
A_- = -i (\mathbb{I}-\l_1 D_2^T \l_2 D_1^T)^{-1}\l_1 (J_{2-} + D^T_2\l_2 J_{1-})
\end{split}
\ee
and that
\be
\label{abm}
\begin{split}
& B_+ = i (\mathbb{I}-\l_2^T D_2 \l_1^T D_1)^{-1}\l_2^T (J_{2+} +  D_2\l_1^T J_{1+})\ ,
\\
&
B_- = -i (\mathbb{I}-\l_2 D_1^T \l_1 D_2^T)^{-1} \l_2 (J_{1-} +  D^{T}_1\l_1 J_{2-})\ ,
\end{split}
\ee
where we have used the definitions \eqn{hg3} with the indices $1$ and $2$ indicating
the usage of the group elements $g_1$ and $g_2$ in the definitions.
Substituting back into the action we obtain the $\s$-model action written in matrix
notation as
\be
\boxed{\
\begin{split}
&  S_{k,\l_1,\l_2}(g_1,g_2) = S_{k}(g_1) + S_{k}(g_2)
\\
&\qq\quad + {k\ov \pi} \int  d^2\s  \left(\!\! \begin{array}{cc}
    J_{1+}\! &\! J_{2+} \end{array}\!\!  \right)
\left(  \begin{array}{cc}
     \L_{21}\l_1 D_2^T\l_2 &   \L_{21}\l_1 \\
     \L_{12}\l_2 & \L_{12} \l_2 D_1^T\l_1\\
  \end{array} \right)
  \left(\!\! \begin{array}{c}
    J_{1-} \\ J_{2-} \end{array}\!\!\! \right)
\end{split}
\ }\ ,
\label{defactigen}
\ee
where we have defined the matrices
\be
\L_{12}= (\mathbb{I} - \l_2 D_1^T \l_1 D_2^T)^{-1}\ ,\qq
\L_{21}= (\mathbb{I} - \l_1 D_2^T \l_2 D_1^T)^{-1}\ .
\ee
This action has by construction the interchanging symmetry of
the two original models, i.e. indices $1$ and $2$.
More importantly, it also has a remarkable duality-type symmetry\footnote{
Combining with the model interchanging symmetry this is equivalent to
\be
 k \to -k \ , \quad \l_1 \to \l_2^{-1} \ ,\quad  \l_2 \to \l_1^{-1}\ ,
 \quad g_1\to g_1^{-1}\quad g_2\to g_2^{-1}\  .
\ee
}
\be
 k \to -k \ , \quad \l_1 \to  \l_1^{-1} \ ,\quad  \l_2 \to  \l_2^{-1}\ ,
 \quad g_1\to g_2^{-1}\quad g_2\to g_1^{-1}\  .
\label{symmdual}
\ee
which is reminiscent of \eqn{duallli}. To prove that we have used that under \eqn{symmdual}
\be
\begin{split}
&
D_1\to D_2^T\ ,\qquad J_{1+}\to - D_2^T J_{2+}\ ,\qq J_{1-}\to - D_2 J_{2-}\  ,
\\
&
D_1\to D_2^T\ ,\qquad  J_{2+}\to - D_1^T J_{1+}\ , \qq J_{2-}\to -  D_1 J_{1-}\  .
\end{split}
\ee
Depending on the form of the $\l_i's$ the action \eqn{defactigen} may have global isometries.
For $\l_i$'s proportional to the identity it has the following global symmetry
\be
g_1\to \L_L^{-1} g_1 \L_R \ ,\qq   g_2\to \L_R^{-1} g_2 \L_L\ , \qq \L_L,\L_R\in G\ .
\ee
For other choices this symmetry is broken partially or all together.

\no
For small elements of the matrices $\l_i$'s the action \eqn{defactigen} becomes
\be
\label{expan1}
S_{k,\l_1,\l_2}(g_1,g_2) = S_{k}(g_1) + S_{k}(g_2) + {k\ov \pi}
\int d^2\s\ (\l_1^{ab} J^a_{1+} J^b_{2-} + \l_2^{ab} J^a_{2+} J^b_{1-}) + \cdots \ .
\ee
thus representing a current-current deformation of the original WZW actions. However,
unlike the $\l$-deformed action \eqn{laact} these current bilinears belong to different
WZW models.\footnote{It is also distinct from the case of \eqn{laact} for the $\l$-deformation
of the direct product $G_{k_1}\times G_{k_2}$ WZW model. For instance in that case the analog
of \eqn{expan1} is
\be
\label{expan2}
S_{k_1,k_2,\l}(g_1,g_2) = S_{k_1}(g_1) + S_{k_2}(g_2) + {k\ov \pi}
\int d^2\s\ \l_{ab} (s_1 J^a_{1+} + s_2 J^a_{2+})(s_1 J^b_{1-} + s_2 J^b_{2-}) + \cdots \ ,
\ee
where $k=k_1+k_2$ and $s_i=k_1/k$, $i=1,2$.
}
Curiously, if one of the matrices $\l_i$ is identically equal to zero the above expression is
exact in the other matrix. Note that this particular case has been examined before in
\cite{Solovev:1993he,Hull:1995gj}.

\no
Due to the similarity with the well known $\l$-deformations and taking into account that
there are two matrices required for a general deformation, the $\s$-model \eqn{defactigen} could be named as a double $\l$-deformation.

\section{Integrability}

In this section we provide the proof that the $\s$-model action \eqn{defactigen}
is integrable for specific choices of the matrices $\l_1$ and $\l_2$.
In particular, we will show that it is integrable for the choices of the
matrices $\l_1$ and $\l_2$ and only for those, which give rise to integrable $\l$-deformed
models corresponding to \eqn{laact} (for $\l=\l_1$ and $\l=\l_2$, separately).

\subsection{Lax pairs}

First recall that we have chosen to gauge fix $\tilde g_i  = \mathbb{I}$, $i=1,2$
in \eqn{fgje3} which immediately led to \eqn{gaufix}. It is easily seen that the equation of
motion followed by varying the $\tilde g_i$'s are automatically satisfied.
When examining integrability of \eqn{defactigen} it is equivalent and more convenient
to work with the gauged fixed action before integrating out the gauge fields \eqn{gaufix}.
Varying the gauged fixed action \eqn{gaufix} with respect
to $B_\pm$ and $A_\pm$, we find the constraints
\be
\qq D_+ g_1\, g_1^{-1} =  (\l_1^{-T}-1) A_+ \ ,\qq
g_2^{-1} D_- g_2 = - (\l_1^{-1}-1) A_-
\label{dggd}
\ee
and
\be
\qq D_+ g_2\, g_2^{-1} =  (\l_2^{-T}-1) B_+ \ ,\qq
g_1^{-1} D_- g_1 = - (\l_2^{-1}-1) B_- \ ,
\label{dggd2}
\ee
respectively. Varying with respect to $g_1$ and $g_2$ we obtain  that
\be
\label{eqg1g2}
D_ -(D_+ g_1 g_1^{-1})= F_{+-}^{(A)}\ ,\qq D_ -(D_+ g_2 g_2^{-1})
= F_{+-}^{(B)}\ ,
\ee
where
\be
F_{+-}^{(A)}=\del_+ A_- - \del_- A_+ - [A_+,A_-]\ ,\qq
F_{+-}^{(B)}=\del_+ B_- - \del_- B_+ - [B_+,B_-]\ .
\ee
Equivalently, these can be written as
\be
D_+(g_1^{-1}D_- g_1)= F_{+-}^{(B)}\ ,\qq
D_+(g_2^{-1}D_- g_2)= F_{+-}^{(A)}\ .
\label{eqg1g22}
\ee
The various covariant derivatives are defined according to the transformation properties of
the object they act on. For instance, $D_\pm g_1= \del_\pm g_1 -A_\pm g_1 + g_1B_\pm$ which
involves both the $A_\pm$ and the $B_\pm$ gauge fields, but
$D_- (D_+ g_1  g_1^{-1})= \del_-(D_+ g_1  g_1^{-1}) -[A_-,(D_+ g_1  g_1^{-1})]$
involves only the $A_\pm$.
Substituting the constraint equations into \eqn{eqg1g2} and \eqn{eqg1g22} we obtain after some algebra
that\footnote{
In components
\begin{equation*}
M A_+=M_{bc}A_+^c\,t_b\,,\qq [M A_+,A_-]=f_{bcd}M_{ce}A_+^e A^d_-\,t_b\,,
\end{equation*}
where $M$ is an arbitrary square matrix.
}
\be
\begin{split}
\label{eomAinitial1}
&\del_+ A_- - \del_- (\l_1^{-T} A_+) = [\l_1^{-T} A_+,A_-]\ ,
\\
& \del_+(\l_1^{-1}A_-)-\del_-A_+=[A_+,\l_1^{-1}A_-]\
\end{split}
\ee
and
\be
\begin{split}
\label{eomAinitial2}
&  \del_+ B_-  - \del_- (\l_2^{-T} B_+) = [\l_2^{-T} B_+,B_-]\ ,
\\
&
\del_+ (\l_2^{-1}B_-)- \del_-B_+=[B_+,\l_2^{-1}B_-]\ .
\end{split}
\ee
Hence the equations of motion split into two identical sets which are seemingly decoupled
even though the fields $A_\pm$ and $B_\pm$ depend on $(g_1,\l_1)$ as well as
on $(g_2,\l_2)$.
Moreover, each set is the same one that one would have obtained had
he performed the corresponding analysis for the $\l$-deformed action \eqn{laact}
(before integrating out the gauge fields, see subsection 2.2 of \cite{Sfetsos:2014lla})
using either $\l=\l_1$ or $\l=\l_2$.
Hence, all choices for matrices known to give rise to integrability
for the $\l$-deformed models provide integrable models here as well.
There is a caveat however. One should show that the conserved
changes obtained by these sets of integrable models are in involution. This task will
performed successfully in the next subsection.

\no
Below we list the  known cases where the matrices
 $\l_1$ and $\l_2$ give rise to
integrability for the system  \eqn{eomAinitial1} and \eqn{eomAinitial2}, respectively.
We also present the Lax pairs satisfying
\be
\label{Lax}
\del_+\cL^{(i)}_- -\del_-\cL^{(i)}_+=[\cL^{(i)}_+,\cL^{(i)}_-]\ ,\qq i=1,2\ ,
\ee
where $\cL^{(i)}_\pm(\tau,\sigma;\zeta_i)$ should depend on a spectral
parameter $\zeta_i\in\mathbb{C}$. Since the two sets of equations are decoupled and for notational convenience
we drop the subscript from $\l_i$ and $\zeta_i$. We will also concentrate on the system \eqn{eomAinitial1}
and the fields $A_\pm$.

\no
{\bf Case 1:} The $\l$-deformation of the isotropic PCM case with $\l^{ab}= \l \d_{ab}$. Then
\be
\del_\pm A_\mp = \pm {1\ov 1+\l} [A_+,A_-]\ ,\qq
\cL_\pm  = {2\ov 1+\l} {\zeta \ov \zeta \mp 1} A_\pm \ ,\quad \zeta\in\mathbb{C}\ .
\label{Laxpairs}
\ee
Using the first relation we easily establish \eqn{Lax}. For this case integrability was shown in \cite{Sfetsos:2013wia} and also in \cite{Hollowood:2014rla}. 

\no
{\bf Case 2:} The $\l$-deformation of the $G/H$ symmetric coset space.\\ In this case
$\l= \diag(\mathbb{I}_{\dim\!H} , \l \mathbb{I}_{\dim\! G/H})$
(of course $\l$ on the right hand side is a single parameter).
In this case it is convenient to split the matrix in its subgroup and coset components as
\be
 A_\pm = A^h_\pm +  A^{g/h}_\pm \ ,\qq A^h_\pm\in \cL(H)\ ,\qq A^{g/h}_\pm \in \cL(G/H)\ ,
\label{sdjfhk}
\ee
where by $\cL (H)$ and $\cL(G/H)$ we denoted the generator of $\cL(G)$ having subgroup and
coset indices, respectively. If the $G/H$ is a symmetric space,
then the system \eqn{eomAinitial1} becomes
\be
\label{eom.symmetric}
\begin{split}
&\del_+ A^h_- - \del_-A^h_+=[A^h_+,A^h_-]+\frac{1}{\l}[A^{g/h}_+,A^{g/h}_-]\,,\\
&\del_\pm A^{g/h}_\mp=-[A^{g/h}_\mp,A^h_\pm]\ .
\end{split}
\ee
Then, the Lax pair is given by \cite{Hollowood:2014rla}
\be
\cL_\pm=A^h_\pm+ {\zeta^{\pm 1}\ov \sqrt{\l}}\,A^{g/h}_\pm\ , \qquad \zeta\in\mathbb{C} \ .
\ee
It can be readily checked that then \eqn{Lax} is satisfied.

\no
{\bf Case 3:} The special case where $G=SU(2)$ and general diagonal matrix given by\\
$\l= \diag(\l_1,\l_2,\l_3)$.
%(there should be no confusion form using the same notation for the matrix and its $(11)$ element).
Integrability was established in \cite{Sfetsos:2014lla} and below we just present the result.
In a basis in which $t_a=-i \s_a/\sqrt{2}$, where $\s_a$ are the Pauli matrices
we have that
\be
\label{eomsu2}
\del_\pm A_\mp^1 = {\sqrt{2}\,\l_1\ov (1-\l_1^2)\l_2 \l_3} \left[(\l_2-\l_1\l_3)A_\pm^2 A_\mp^3 -
(\l_3-\l_1\l_2) A_\pm^3 A_\mp^2\right]\
\ee
and cyclic in $1,2$ and $3$.
Then the Lax pair $\cL_{\pm} = \cL^a_{\pm}t^a$ has
\be
\label{Lax.su2}
\cL^{a}_{\pm}= \sqrt{z_\pm(\zeta) + c_a^2}\  X^a_\pm \ ({\rm no\ sum\ over}\ a)\ ,\quad z_\pm \in \mathbb{C}\ ,
\quad a=1,2,3\ ,
\ee
where
\be
X^1_\pm = {A_\pm^1\ov \l_1 \sqrt{(1-\l_2^2)(1-\l_3^2)} } \ ,\qq c_1 = \l_1 - \l_2\l_3\
\ee
and cyclic in $1,2$ and $3$. The constants $z_\pm$ obey the condition
\be
(z_+z_- - c_1^2 c_2^2 - c_2^2 c_3^2 - c_3^2 c_1^2)^2 = 4 c_1^2  c_2^2 c_3^2(z_+ +z_- + c_1^2+c_2^2 +c_3^2)\ ,
\ee
which when solved introduces the spectral parameter $\zeta$. When $\l_1=\l_2=\l_3$
the results of this case reduce to those of case 1 for $G=SU(2)$.

\no
{\bf Case 4:} The $\l$-deformation of the YB model \cite{Sfetsos:2015nya}.  Then the matrix $\l$ is given by
\eqn{defl} with
\be
E = \frac{1}{\tilde t} (\mathbb{I} - \tilde\eta {\cal R})^{-1} \ ,
\ee
which is a two-parameter deformation.
The matrix $\cR$ satisfies the modified Yang--Baxter equation written explicitly as
\be
\label{mYB1}
c^2\,f_{abc} + \cR_{ad} \cR_{be} f_{dec} +\cR_{bd} \cR_{ce} f_{dea} + \cR_{cd} \cR_{ae} f_{deb}=0\ ,
\ee
with $c^2=1,-1,0$. For compact groups it is necessary to have that $c^2=-1$ for
a solution of this equation to exist. Referring for
details to \cite{Sfetsos:2015nya} we present briefly the results.
Defining that
\be
\widetilde A_\pm=(\mathbb{I}  \pm\tilde\eta\,\cR)^{-1}\,A_\pm\ ,
\ee
and that
\be
a=\frac{1+c^2\tilde\eta^2\lambda_0}{1+\lambda_0}\ , \qq  \lambda_0=\frac{k\,\tilde t}{1+k\,\tilde t}\,.
\ee
we have from \eqn{eomAinitial1} that
\be
\pm\del_\pm \widetilde A_\mp=\tilde\eta[\cR \widetilde A_\pm,\widetilde A_\mp]+a\,[\widetilde A_+,\widetilde A_-]\ .
\ee
The Lax par is
\be
\label{LaxYB}
\begin{split}
&{\cal L}_\pm=(\a_\pm\mathbb{I}\pm\tilde \eta\,{\cal R})(\mathbb{I}\pm\tilde\eta\,{\cal R})^{-1}\,A_\pm\,,\\
&\a_\pm=\a_1+\a_2\,\frac{\zeta}{\zeta\mp1}\,,\quad \zeta\in\mathbb{C}\,,\\
&\a_1=a-\sqrt{a^2-c^2\tilde\eta^2}\,,\quad \a_2=2\sqrt{a^2-c^2\tilde\eta^2}\, .
\end{split}
\ee
There is coset version of this case \cite{Sfetsos:2015nya} but will refrain
from presenting any details here.

\no
Evidently an identical list holds for the system \eqn{eomAinitial2} for the $B_\pm$ gauge fields.
Therefore, for a given general group $G$ we have, by combining different
choices for the matrices $\l_1$ and $\l_2$, three independent integrable models corresponding
to the action \eqn{defactigen}. For a general symmetric
coset space $G/H$ we have also an additional integrable model.

\subsection{Canonical treatment and charges in involution}

The above Lax pairs $\cL^{(i)}_\pm$ are enough to prove integrability provided that the
corresponding charges are in involution.
In order to verify this  we will calculate the Dirac bracket of the
gauge fields $A_{\pm}$ and $B_{\pm}$ and show that it is zero implying the independence of the
infinite charges generated by $\cL^{(1)}_\pm$ from those generated by $\cL^{(2)}_\pm$.
We will follow the corresponding treatment of
 \cite{Bowcock} for the case of gauged WZW models.

\no
We parametrize the group elements $g_i$ and $g_2$ by coordinates $x^\m_1$ and $x_2^\m$, with $\m=1,2,\dots , \dim G$.
Then we define the matrices
\be
\begin{split}
& L^a_{1\m} = -i {\rm Tr}(t^a g_1^{-1}\del_\m g_1)\ ,\qq
R^a_{1\m} = -i {\rm Tr}(t^a \del_\m g_1 g_1^{-1}) = D_1^{ab}L^a_{1\m}\ ,
\\
&  L^a_{2\m} = -i {\rm Tr}(t^a g_2^{-1}\del_\m g_2)\ ,\qq
R^a_{2\m} = -i {\rm Tr}(t^a \del_\m g_2 g_2^{-1}) = D_2^{ab}L^a_{2\m}\ .
\end{split}
\ee
Then the action \eqn{gaufix} becomes
\be
\begin{split}
&  S_{k}(x_1,x_2, A_\pm, B_\pm)  = {k\ov 4\pi}
 \int d^2\s\   \Big(\ha R^a_{1\m} R^a_{1\n}(\dot x_1^\m \dot x_1^\n
- x_1^{\prime\m}   x_1^{\prime\n}) +
\l_{\m\n}^{(1)}\dot x_1^\m x_1^{\prime \n}
\\
&\qq + R^a_{2\m} R^a_{2\n}(\dot x_2^\m \dot x_2^\n
- x_2^{\prime\m}   x_2^{\prime\n} )+
\l_{\m\n}^{(2)}\dot x_2^\m x_2^{\prime \n}
\\
&\qq +  2i A_-^a R^a_{1\m} (\dot x_1^\m+x_1^{\prime \m})
 -2 i B_+^a  L^a_{1\m}(\dot x_1^\m - x_1^{\prime\m})
 +4 B_+^a D_1^{ba}A_-^b  - 4  A_+^a \l_{1\,ab}^{-1}A_-^b
\label{Scan}
 \\
 &\qq +  2i B_-^a R^a_{2\m} (\dot x_2^\m+x_2^{\prime \m})
 -2 i A_+^a  L^a_{2\m}(\dot x_2^\m - x_2^{\prime\m})
 +4 A_+^a D_2^{ba}B_-^b  - 4  B_+^a  \l_{2\,ab}^{-1}B_-^b \Big)\ ,
 \end{split}
\ee
where $\l^{(1)}_{\m\n}$ and $\l^{(2)}_{\m\n}$ are the antisymmetric couplings of the two WZW actions.
From this action we deduce the canonical momenta as
\be
\label{momenta}
\begin{split}
&\p_\m^{(1)}={\partial {\cal L}\ov \partial \dot x_1^\m}=
{k\ov 4\pi} \big( R^a_{1\m} R^a_{1\n}\dot x_1^\n + \l_{\m\n}^{(1)} x_1^{\prime \n}
+ 2 i A_-^a R^a_{1\m} - 2 i B_+^a L^a_{1\m}\big)\ ,
 \\
 &\p_\m^{(2)}={\partial {\cal L}\ov \partial \dot x_2^\m}=
{k\ov 4\pi} \big( R^a_{2\m} R^a_{2\n}\dot x_2^\n + \l_{\m\n}^{(2)} x_2^{\prime \n}
+  2 i B_-^a R^a_{2\m} - 2 i A_+^a L^a_{2\m}\big)\ ,
\\
 & P_\pm^a= {\partial {\cal L}\ov \partial \dot A_\pm^a}=0\ , \qq
  Q_\pm^a= {\partial {\cal L}\ov \partial \dot B_\pm ^a}=0\ .
\end{split}
\ee
These obey the usual Poisson bracket relations.
The vanishing of the canonical momenta conjugate to the gauge fields $A^a_{\pm}$ and $B^a_{\pm}$ imply the existence of constraints
in the Hamiltonian formalism and one should invoke Dirac's procedure in order to deal with them. To this end, we derive
the Hamiltonian of our system, in the usual way. It takes the form
\be
\begin{split}
&  H_0  ={k\ov 4\pi}
 \int d\s\   \Big(\ha R^a_{1\m} R^a_{1\n}(\dot x_1^\m \dot x_1^\n
+ x_1^{\prime\m}   x_1^{\prime\n})
 +\ha  R^a_{2\m} R^a_{2\n}(\dot x_2^\m \dot x_2^\n +  x_2^{\prime\m}   x_2^{\prime\n} )
\\
&\qq -  2i A_-^a R^a_{1\m} x_1^{\prime \m}
 -2 i B_+^a  L^a_{1\m}  x_1^{\prime\m}
 - 4 B_+^a D_1^{ba}A_-^b  + 4  A_+^a  \l_{1\,ab}^{-1}A_-^b
\label{H0}
 \\
 &\qq -  2i B_-^a R^a_{2\m} x_2^{\prime \m}
 -2 i A_+^a  L^a_{2\m}  x_2^{\prime\m}
 - 4 A_+^a D_2^{ba}B_-^b  + 4  B_+^a  \l_{2\,ab}^{-1}B_-^b \Big)\ .
 \end{split}
\ee
In order to proceed, we define the currents $J_{i\pm}^a$, $i=1,2$ as the same functions of the phase space variables
appearing in the ungauged WZW model, that is
\be
\label{currents}
\begin{split}
&
J^a_{1+}={1 \ov 2} R_1^{a\m}\left({4\pi \ov k}\pi^{(1)}_\m - \l^{(1)}_{\m\n}x_1^{\prime \n}\right)+\ha  R^a_{1\mu} x_1^{\prime \m}
\\
&\qq\qq
= \ha R^a_{1\m} (\dot x_1^\m + x_1^{\prime \m}) + i A_-^a - i D_1^{ab}B_+^b\ ,
\\
& J^a_{1-}={1 \ov 2} L_1^{a\m}\left({4\pi \ov k}\pi^{(1)}_\m - \l^{(1)}_{\m\n}x_1^{\prime \n}\right)-\ha  L^a_{1\mu} x_1^{\prime \m}
\\
& \qq\qq
= \ha L^a_{1\m} (\dot x_1^\m - x_1^{\prime \m}) - i B_+^a + i D_1^{ba}A_-^b\  ,
\end{split}
\ee
where $ R^{a\m}_1$ and $ L^{a\m}_1$ are the inverses of $R^a_{1\m}$ and $L^a_{1\m}$, respectively.
In addition, we have used the identity $R^a_{1\m}R^a_{1\n}=L^a_{1\m}L^a_{1\n}$.
The expressions for the currents $J^a_{2\pm}$ can be obtained from \eqref{currents} after substituting the index $1$ by $2$
 and the gauge field $A_{\pm}$ with $B_{\pm}$.
Note that the $J^a$'s in \eqn{currents} are not the same as the ones defined in
\eqn{hg3}, but we have used the same symbols so that we restrict the number of
different symbols to a minimum necessary.

\no
In this way the Poisson brackets (PB) of the currents take the same form as in the
ungauged WZW theory since the PB of the canonical variables $x^\m_{1,2}$ and $\pi_\m^{(1,2)}$ remain unchanged.
After some algebra the Hamiltonian \eqref{H0} can be written as
\ba
&&  H_0  = {k\ov 4\pi} \int d\s \   \Big( J^a_{1+}  J^a_{1+} + J^a_{1-}  J^a_{1-}
 + J^a_{2+}  J^a_{2+} + J^a_{2-}  J^a_{2-}
 \nonumber
\\
&&\qq\qq  +4 i (J^a_{1-}B^a_+ - J^a_{1+}A^a_- + J^a_{2-}A^a_+ - J^a_{2+}B^a_-)
\nonumber\\
&& \qq\qq  -2(A^a_+-A^a_-)(A^a_+-A^a_-)-2(B^a_+-B^a_-)(B^a_+-B^a_-)
\label{H1}
\\
&& \qq\qq + 4 A_+^a (\l_{1}^{-1}-\mathbb{I})_{ab}A_-^b  + 4 B_+^a (\l_{2}^{-1}-\mathbb{I})_{ab}B_-^b \Big)\ .
 \nonumber
 \ea
To this Hamiltonian one should add arbitrary linear combinations of the primary constraints appearing in the last line of
\eqref{momenta}.   The resulting Hamiltonian then reads
\be\label{Ht}
H=H_0+\int d \s (c^a_+P^a_+ +c^a_-P^a_-+d^a_+Q^a_+ + d^a_-Q^a_-)\ .
\ee
The time evolution of the primary constraints generates secondary constraints as
\be
\begin{split}
& [P^a_+,H]=  4\big( A^a_+ -  \l_{1\, ab}^{-1}A_-^b - i J^a_{2-}\big)\approx 0\ ,
\\ \label{sec1}
& [P^a_-,H]= 4\big(A^a_- -\l_{1\, ab}^{-T} A^b_+ +  i J^a_{1+}\big)\approx 0\ ,
\\
& [Q^a_+,H]=  4\big(B^a_+ -  \l_{2\, ab}^{-1} B^b_- - i J^a_{1-} \big)\approx 0\ ,
\\
& [Q^a_-,H]= 4\big( B^a_- -\l_{2\, ab}^{-T} B^b_+ + i J^a_{2+}\big)\approx 0 \ .
\end{split}
 \ee
In order to see if there are any other constraints generated one should find the time derivatives of the
secondary constraints above and set them weakly to zero.  For instance, it is straightforward to check that computing
$[A^a_+ -  \l_{1\, ab}^{-1}A_-^b - i J^a_{2-},H]$ gives an expression of $c^a_{\pm},  d^a_{\pm}$ and similarly for the other three secondary constraints. These equations should be solved to
obtain $ c^a_{\pm}$ and  $d^a_{\pm}$
as functions of the dynamical variables which should then be substituted back in \eqref{Ht} to get the final expression for the Hamiltonian
of the system. We conclude that no additional constraints are generated leaving us with the following eight constraints, all of which are second class. This is due to the fact that our action has no residual gauge symmetry after we have gauge fixed the group elements of the
PCMs to one, i.e. $\tilde g_1=\mathbb{I}=\tilde g_2$. In contradistinction, in the case of the gauged WZW model one encounters both first and second
class constraints \cite{Bowcock}. To summarize, we have the following eight constraints, primary and secondary
\be
\begin{split}
&\chi_1^a=P^a_+\approx 0\ ,
\\
&\chi_2^a=P^a_-\approx 0\ ,
\\
& \chi_3^a=A^a_+ -  \l_{1\, ab}^{-1}A_-^b - i J^a_{2-}\approx 0\ ,
\\ \label{conA}
& \chi_4^a= A^a_- -\l_{1\, ab}^{-T} A^b_+ +  i J^a_{1+}\approx 0 \ ,
\end{split}
 \ee
and
\be
\begin{split}
&\chi_5^a=Q^a_+\approx 0 \ ,
\\
&\chi_6^a=Q^a_-\approx 0\ ,
\\
& \chi_7^a=B^a_+ -  \l_{2\, ab}^{-1} B^b_- - i J^a_{1-}\approx 0\ ,
\\ \label{conB}
& \chi_8^a= B^a_- -\l_{2\, ab}^{-T} B^b_+ + i J^a_{2+}\approx 0\ .
\end{split}
 \ee
Notice that all constraints of \eqref{conA} have zero PBs with any of the constraints of \eqref{conB}.
\begin{comment}
This is so because of the  following PBs
\be
\begin{split}
&[J^i_{1-}(\s),J^j_{1+}(\s')]= 0,\,\,\,[J^i_{2-}(\s),J^j_{2+}(\s')]= 0,\,\,\,[A^i_{\pm}(\s),A^j_{\pm}(\s')]=0,\,\,\,[J^i_{1\pm}(\s),A^j_{\pm}(\s')]= 0,\,\,\,
\\ \label{PB0}
&[J^i_{2\pm}(\s),A^j_{\pm}(\s')]= 0,\,\,\, [A^i_{\pm}(\s),B^j_{\pm}(\s')]=0,\,\,\,[B^i_{\pm}(\s),B^j_{\pm}(\s')]=0.\,\,\,[J^i_{1\pm}(\s),B^j_{\pm}(\s')]= 0\\
& [J^i_{2\pm}(\s),B^j_{\pm}(\s')]= 0.
\end{split}
 \ee
\end{comment}
The next step is to define the antisymmetric matrix
\be
\begin{split}
C_{\a\b}=[\chi_{\a},\chi_{\b}] =
\left(  \begin{array}{cc}
   (C _{1})_{ab} &  {\bf 0}_{4\times 4} \\
     {\bf 0}_{4\times 4} &  (C _{2})_{ab}\\
  \end{array} \right)\ ,  \quad a,b=1,\dots,4\ ,
%\quad \a,\b=a+4,b+4\ ,
 \end{split}
\label{Cmatrix}
\ee
whose precise form is given below.
The Dirac bracket (DB) is defined as
\be
\label{DiracB}
[\zeta, \eta]_{DB}=[\zeta, \eta]-[\zeta, \chi_{\a}] (C^{-1})^{\a\b}[\chi_{\b}, \eta] \ .
\ee
The time evolution of any quantity $\zeta$ is now given by $\dot \zeta\approx [\zeta, H]_{DB}$.
Due to the block diagonal structure of the matrix \eqn{Cmatrix} and of its inverse
there will be no mixing between the two blocks of constraints.

Hence, we conclude that the infinite charges obtained from the
Lax pairs corresponding to the systems \eqn{eomAinitial1} and \eqn{eomAinitial2}
are in involution.
Since the matrix $C^{-1}$ is block diagonal it is straightforward to deduce that
\be\label{independence}
[A^a_{\pm}(\tau,\s_1),B^b_{\pm}(\tau,\s')]_{\rm DB}=0\ ,
\qq [\cL^{(1)}_\pm (\tau,\s_1,\zeta_1),\cL^{(2)}_\pm(\tau,\s',\zeta_2)]_{\rm DB}
=0\ ,
\ee
for all signs independently and for all values of the
spectral parameters $\zeta_1$ and $\zeta_2$.
This fact is not at all obvious from the expressions \eqn{abp}  and \eqn{abm} and since these involve both groups
elements $g_1$ and $g_2$.
Hence, all the charges obtained from the monodromy matrix corresponding to $\cL^{(1)}$
are in involution with those obtained from
the monodromy matrix corresponding to  $\cL^{(2)}$,
proving thus the integrability of the theory \eqn{defactigen}.

\no
We now provide for completeness the expressions for the matrix of the constraints $C_{\a\b}$. To compute them we use the canonical PB for $A_{\pm},B_{\pm}$ and their conjugate momenta $P_{\pm},Q_{\pm}$,
as well as for the currents $J^a_{1\pm}$ and $J^a_{2\pm}$, namely that
\be
\begin{split}
& [J^a_{1\pm}(\s_1),J^b_{1\pm}(\s_2)]
= -{2\pi\ov k} f_{abc}J_{1\pm} ^c(\s_2) \d_{12} \pm {2\pi \ov k} \d_{ab} \d'_{12}\ ,
\\
& [J^a_{2\pm}(\s_1),J^b_{2\pm}(\s_2)]= -{2\pi\ov k} f_{abc}J_{2\pm} ^c(\s_2) \d_{12}
\pm {2\pi \ov k} \d_{ab} \d'_{12}\ ,
\\
& [J^a_{1+}(\s_1),J^b_{1-}(\s_2)]=0\ ,\quad [J^a_{2+}(\s_1),J^b_{2-}(\s_2)]=0\ ,
\\
&[J^a_{1\pm}(\s_1),J^b_{2\mp}(\s_2)]=0\ ,\quad [J^a_{1\pm}(\s_1),J^b_{2\pm}(\s_2)]=0\ ,
\\
&[A^a_{\pm}(\s_1),P^b_{\pm}(\s_2)]=\d^{ab} \d(\s_{12})\ ,\,\,\,[B^a_{\pm}(\s_1),Q^b_{\pm}(\s_2)]=\d^{ab} \d(\s_{12})\ ,
\end{split}
\ee
with all other PBs being zero.
Then we find that
\be
%\boxed{\
%\begin{split}
C_{1}=
\left(  \begin{array}{cccc}
   0 & 0 &  -\d^{ab}_{12} &  (\l_1^{-T})^{ab}\d_{12}\\
     0 &  0 &  (\l_1^{-1})^{ab}\d_{12} &  -\d^{ab}_{12}\\
     \d^{ab}_{12}  &  -  (\l_1^{-1})^{ab}\d_{12}& C^{ab}_{33} & 0\\
    - (\l_1^{-T})^{ab}\d_{12} &  \d^{ab}_{12} & 0 & C^{ab}_{44}\\
  \end{array} \right),\,
 %\end{split}
%\ }
\label{C1matrix}
\ee
where $ \d^{ab}_{12}= \d_{ab} \d(\s_{12})$
and
\be
\begin{split}
&C_{33}^{ab}(\s_1,\s_2)={2 \pi \ov k} \d_{ab}\d'(\s_{12})-{2 \pi \ov k} i f_{abc}\d(\s_{12})\Big( A_+^c(\s_2)-  \l_{1\, cd}^{-1}A_-^d(\s_2) \Big)\\
&C_{44}^{ab}(\s_1,\s_2)=-{2 \pi \ov k} \d_{ab}\d'(\s_{12})+{2 \pi \ov k} i f_{abc}\d(\s_{12})\Big( A_-^c(\s_2)-  \l_{1\, cd}^{-T}A_+^d(\s_2) \Big).
\end{split}
\ee
It is straightforward to evaluate the inverse of \eqref{C1matrix} needed in the definition of the Dirac brackets. We find that
\be
\begin{split}
& C_{1}^{-1}= \\
 & \left(  \begin{array}{cccc}
  F_1^{ab} &F_2^{ab} &  (\mathbb{I}-\l_{1}^{-1}\l_{1}^{-T})^{-1}_{ab}\d_{12} &   (\l_1-\l_{1}^{-T})^{-1}_{ab}\d_{12} \\
   F_3^{ab}&  F_4^{ab} &   (\l_{1}^{T}-\l_1^{-1})^{-1}_{ab}\d_{12} & (\mathbb{I}-\l_{1}^{-T}\l_{1}^{-1})^{-1}_{ab}\d_{12}\\
     (\l_{1}^{-T}\l_{1}^{-1}-\mathbb{I})^{-1}_{ab}\d_{12}  &  (\l_1^{-1}-\l_{1}^{T})^{-1}_{ab}\d_{12} & 0& 0\\
    (\l_{1}^{-T}-\l_1)^{-1}_{ab}\d_{12} & (\l_{1}^{-1}\l_{1}^{-T}-\mathbb{I})^{-1}_{ab}\d_{12} & 0 & 0\\
  \end{array} \right)\  ,
 \label{C1matrix}
 \end{split}
\ee
where
\be
\begin{split}
& F_1^{ab}= (\l_{1}^{-1}\l_{1}^{-T}-\mathbb{I})^{-1}_{ac}C_{33}^{cd}(\s_1,\s_2)
(\l_{1}^{-T}\l_{1}^{-1}-\mathbb{I})^{-1}_{db}
\\
&\qq\qq\qq\qq\qq\qq
+ (\l_{1}^{-T}-\l_1)^{-1}_{ac}C_{44}^{cd}(\s_1,\s_2) (\l_{1}^{-T}-\l_1)^{-1}_{db}\ ,
\\
 & F_2^{ab}= (\l_{1}^{-1}\l_{1}^{-T}-\mathbb{I})^{-1}_{ac}C_{33}^{cd}(\s_1,\s_2)
(\l_1^{-1}-\l_{1}^{T})^{-1}_{db}
\\
& \qq\qq\qq\qq\qq\qq
+ (\l_{1}^{-T}-\l_1)^{-1}_{ac}C_{44}^{cd}(\s_1,\s_2)
(\l_{1}^{-1}\l_{1}^{-T}-\mathbb{I})^{-1}_{db} \ ,
\\
& F_3^{ab}= (\l_{1}^{-1}-\l_1^T)^{-1}_{ac}C_{33}^{cd}(\s_1,\s_2)
(\l_{1}^{-T}\l_{1}^{-1}-\mathbb{I})^{-1}_{db}
\\
& \qq\qq\qq\qq\qq\qq
+ (\l_{1}^{-T}\l_{1}^{-1}-\mathbb{I})^{-1}_{ac}C_{44}^{cd}(\s_1,\s_2) (\l_{1}^{-T}-\l_1)^{-1}_{db} \\
& F_4^{ab}= (\l_{1}^{-1}-\l_1^{T})^{-1}_{ac}C_{33}^{cd}
(\s_1,\s_2)(\l_1^{-1}-\l_{1}^{T})^{-1}_{db}
\\
& \qq\qq\qq\qq\qq\qq
+ (\l_{1}^{-T}\l_{1}^{-1}-\mathbb{I})^{-1}_{ac}C_{44}^{cd}(\s_1,\s_2)
(\l_{1}^{-1}\l_{1}^{-T}-\mathbb{I})^{-1}_{db} \ .
\label{Fs}
 \end{split}
\ee  % (\l_1^{-1}-\l_{1}^{T})^{-1}_{ac}
Finally, the matrix $C_2$ and its inverse $C_2^{-1}$ can be obtained from  $C_1$ and $C_1^{-1}$ after the exchange of
$A_{\pm}\leftrightarrow B_{\pm}$ and of the index $1$ with the index $2$.

\no
Consider now, in cases that this is possible, the decomposition of a semi-simple group G to a semi-simple subgroup
$H$ and a symmetric coset $G/H$. This case can be addressed by taking the coupling matrices  $\l_1$ and $\l_2$ to have the following
non-zero elements, $(\l_i)_H^{ab}=\d^{ab},\,\,i=1,2$ and
$(\l_i)_{G/H}^{\a\b}=(\l_i)_{G/H} \d^{\a\b},\,\,i=1,2$, where now the Latin (Greek) indices
take values in the subgroup $H$ (coset $G/H$).
The Hamiltonian analysis of this case remains the same up to \eqref{Cmatrix}
with the only difference that one should now distinguish
the indices of the constraint matrix belonging to the subgroup $H$ from those belonging
to the coset $G/H$. Nevertheless, the matrix $C$ remains diagonal
implying that the gauge fields $A_{\pm}$ and $B_{\pm}$
from which the two Lax pairs are built still commute.
However, there is a major difference. Because the  zero components of the  gauge
fields $A_{0}^a$ and $B_{0}^a$ do not appear in any of the constraints when the index
$a \in H$ the corresponding
canonical momentum $P_{0}^a$ and $Q_{0}^a$ become first class constraints.
This is in accordance with the fact that the coset theory has a
local gauge symmetry with symmetry group $H$ and implies that the matrices
$C_1$ and $C_2$ have zero determinants. The situation encountered here is
identical to that in \cite{Bowcock} for the usual gauged WZW models and one can follow the usual procedure in order
to treat this case canonically as well.

\section{An example}

In this section we present an example of our construction for the coset ${SU(2) \times SU(2) \ov U(1) \times U(1)} $.
We parametrize an $SU(2)$ group element in terms of Euler angles as
\be
 g=e^{{i \ov 2} \phi\s_3}e^{{i \ov 2} \theta\s_2}e^{{i \ov 2} \psi\s_3}=
\left(
   \begin{array}{cc}
     \cos{{\theta \ov 2}} \ e^{{i \ov 2} (\phi+\psi)}  & \sin{{\theta \ov 2}} \ e^{{i \ov 2} (\phi-\psi)} \\
     -\sin{{\theta \ov 2}} \ e^{-{i \ov 2} (\phi-\psi)} &  \cos{{\theta \ov 2}} \ e^{-{i \ov 2} (\phi+\psi)}  \\
   \end{array}
 \right).
\label{su2g2}
\ee
As usual, $\s_i$, $i=1,2,3$ are the Pauli matrices.
In these coordinates the WZW action reads
\be
\begin{split}
& S_k(g) =  {k\ov 4\pi} \int d^2\s\, \Big[\del_+ \th \del_- \th+\del_+ \phi \del_- \phi+\del_+ \psi \del_- \psi
+2 \cos\th \del_+ \phi \del_- \psi  \Big] \ .
\end{split}
\label{jhk244}
\ee
In our case we have two $SU(2)$ group elements $g_i$, $i=1,2$ and therefore we have
six coordinates $(\th_1,\phi_1,\psi_1)$ and $(\th_2,\phi_2,\psi_2)$.
We will choose to gauge the $U(1) \times U(1)$ symmetry which allows to set $\phi_1=0=\psi_1$.
In order to simplify the notation, we will rename $\phi_2$ and $\psi_2$ to $\phi$ and $\psi$, respectively.

\no
The expressions for the metric and the antisymmetric tensor fields become significantly simpler if one of the couplings
is set to zero, i.e. $\l_2=0$, the reason being that, as we will see, the background develops a $U(1)$ isometry.
We rename $\l_1$ to $\l$ and define for convenience the
function
\be
\D=1 - \cos\th_1 \cos\th_2 + \l \cos\phi \sin\th_1 \sin\th_2\  .
\label{deldef}
\ee
Then, the metric components are (it is more convenient to present the expressions of
most of the metric and antisymmetric tensor components below, multiplied by the function $\D$ defined in \eqn{deldef})
\be
\begin{split}
&G_{\th_1\th_1}  =1\ , \qq G_{\th_2\th_2}  = 1\ ,
\\
&
\D G_{\phi \phi}  = 1 + \cos\th_1 \cos\th_2 + \l \cos\phi \sin\th_1 \sin\th_2 \ ,
\\
&\D G_{\psi \psi} =1 + \cos\th_1 \cos\th_2 - \l \cos\phi \sin\th_1 \sin\th_2 \ ,
\\
& \D G_{\phi \psi}  = \cos\th_1 + \cos\th_2\ ,
\\
&\D G_{\th_1\th_2} = \l (\cos\phi(1- \cos\th_1 \cos\th_2) + \l \sin\th_1 \sin\th_2) \ ,
\\
&\D G_{\th_1\phi} = \l  \sin\phi  \cos\th_1 \sin\th_2\ ,\qq \D G_{\th_1\psi}  = \l \sin\phi \sin\th_2\ ,
\\
&
\D G_{\th_2\phi} = \l \sin\phi \sin\th_1 \cos\th_2  \ ,\qq \D G_{\th_2\psi}  = \l \sin\phi \sin\th_1 \  .
\end{split}
\ee
For the antisymmetric field one obtains
\be
\begin{split}
&\D B_{\th_1\th_2}  = \l \left(\cos\phi (1- \cos\th_1 \cos\th_2 )+ \l \sin\th_1 \sin\th_2\right) \ ,
\\
&
\D B_{\th_1\phi}  = \l \sin\phi \cos\th_1 \sin\th_2 \ ,\qq \D B_{\th_1\psi}  =\l \sin\phi \sin\th_2\ ,
\\
& \D B_{\th_2\phi} = -\l \sin\phi \sin\th_1 \cos\th_2  \ ,\qq \D B_{\th_2\psi}  =-\l \sin\phi \sin\th_1\ ,
\\
 &\D B_{\phi \psi}  = \cos\th_2 - \cos\th_1 \ .
 \end{split}
\ee
The isometry corresponds to shifts of the angular coordinate $\psi$.
It is interesting to look more closely at the CFT background this deformation corresponds to.
Hence, setting furthermore $\l=0$ this corresponds to the following action
\ba
&&  S_k=  {k\ov 4\pi} \int d^2\s\, \Big[\del_+ \th_1 \del_- \th_1+\del_+ \th_2 \del_- \th_2
 + {1+\cos\th_1 \cos\th_2 \ov  1-\cos\th_1 \cos\th_2  }(\del_+ \phi \del_- \phi+\del_+ \psi \del_- \psi)
\nonumber\\
&&
\qq\qq +2{ \cos\th_2 \del_+ \phi \del_- \psi + \cos\th_1 \del_+ \psi \del_- \phi \ov 1-\cos\th_1 \cos\th_2  } \Big] \ ,
\label{paraferm1}
\ea
which has a further isometry corresponding to shifts of $\phi$ as well. It is convenient to
change variables as $\phi=\phi_1+\phi_2$ and  $\psi=\phi_1-\phi_2$. Then the action becomes
\ba
&&  S_k =  {k\ov 4\pi} \int d^2\s\, \Big[\del_+ \th_1 \del_- \th_1+\del_+ \th_2 \del_- \th_2 + {2\ov 1- \cos\th_1 \cos\th_2}
\Big( (1+\cos\th_1)(1+\cos\th_2 ) \del_+ \phi_1\del_- \phi_1
\nonumber\\
&& +  (1-\cos\th_1)(1-\cos\th_2 ) \del_+ \phi_2\del_- \phi_2 + 2 (\cos\th_1-\cos\th_2)
(\del_+\phi_1\del_-\phi_2- \del_+\phi_2\del_-\phi_1)  \Big) \Big] \ .
\label{paraferm12}
\ea
Performing a T-duality along the $\phi_2$-isometry and changing variables as
$\displaystyle \varphi_1= {\phi_2\ov 2}+\phi_1$ and $\displaystyle \varphi_2= {\phi_2\ov 2}-\phi_1$ we obtain that
\be
 S_k =  {k\ov 4\pi} \int d^2\s\, \Big(\del_+ \th_1 \del_- \th_1
+\cot^2{\th_1\ov 2} \del_+\varphi_1  \del_-\varphi_1
+ \del_+ \th_2 \del_- \th_2
+\cot^2{\th_2\ov 2} \del_+\varphi_2  \del_-\varphi_2\Big) \ .
\label{paraferm12}
\ee
This is nothing but the $\s$-model action corresponding to two copies
of the $SU(2)_k/U(1)$ CFT.\footnote{Had we performed an Abelian T-duality along the $\phi$ isometry of \eqref{paraferm1} we
would have obtained  the same action with $\th_2\to \th_1$ and $\th_1\to \pi+ \th_2$. That explains the above
coordinate transformations before the T-duality was performed.
}
Hence, in our example the deformation is on a T-dual of the background for the $SU(2)_k/U(1)\times SU(2)_k/U(1)$ exact CFT background.
We finally mention that we have also worked out the background fields in the general case in which both $\l_1$ and $\l_2$ are different than zero as well
as for the group case $SU(2)_k \times SU(2)_k$. However, we refrain from presenting all details since the various expressions are not very illuminating.
Another example with $SU(2)$ can be constructed by taking $\l_2=0$ (in the full group) and restricting the remaining $\l_1$ so that the gauging
procedure resembles that of \cite{Guadagnini:1987ty} with a $U(1)$ subgroup. At the CFT point the model should be that of \cite{PandoZayas:2000he}.

\section{Conclusions}

We have constructed a new class of $\s-$models for general semisimple groups and provided specific cases for which these are integrable.
This was achieved by gauging asymmetrically two independent WZW models both at level $k$. The asymmetry in the gauging procedure
renders them anomalous when considered separately, but neverthless their combination is anomaly free.
The resulting action \eqn{defactigen} can be thought of as the effective action of two independent WZW models, but unlike the action of
the known $\l$-deformations \eqn{laact}, the perturbation is driven by current bilinears of the different WZW models as in \eqn{expan1}. This is an important difference and as a consequence this new action is not the sum of two individual $\l$-deformed actions.
In the simplest, isotropic case, our models have three independent parameters.
To demonstrate integrability it was necessary to perform a complete Hamiltonian analysis by carefully treating the second class constraints that arise.

\no
Similarly to $\l$-deformations,
we have revealed the non-petrurbative symmetry \eqn{symmdual} in the parameter space of $\l_1$ and $\l_2$, albeit for large levels $k$.
This should be instrumental
in studying the quantum behaviour of these models along the lines of the similar treatment that we have undergone for the
$\l$-deformations. In particular,  one may follow and appropriately modify
the beta-function computations in \cite{Itsios:2014lca,Sfetsos:2014jfa} using gravitational methods,
as well as the computation of the current and anomalous dimensions of primary operators in \cite{Georgiou:2015nka, Georgiou:2016iom} using CFT methods and the non-perturbative symmetry \eqn{duallli}.

\no
In recent work the $\l$-deformations have been extended to the left-right assymetric case by
considering different levels $k_L$ and $k_R$ for the associated current algebras \cite{Georgiou:2016zyo}. In that case there is no
known effective action analog of \eqn{laact} but one may use as a definition of the model the leading order
expression \eqn{expan2kk} where however, the current algebras are at different levels as mentioned. These models have some very attractive features such as the existense of a new fixed point under the renormalization group flow towards the infrared. The beta-function and current operator anomalous dimensions
have be constructed in \cite{Georgiou:2016zyo} for the isotropic case using low order peturbative results obtained by CFT methods
and a generalization of the symmetry \eqn{duallli}.
It should be possible to perform a similar analysics for the new models we have constructed in the present work starting with the
leading order expression \eqn{expan1} (with $\l_i$'s proportional to the identity).

\no
It would be very interesting to embed our models with low target spaces to supergravity, by extending our construction to cases involving non-compact groups. The example provided in section 4 seems to be resonably simple for that purpose.
The analog works of embedding $\l$-deformations to supergravity in \cite{Sfetsos:2014cea,Demulder:2015lva,Borsato:2016zcf,Chervonyi:2016ajp,Chervonyi:2016bfl} should be very helpful in that direction. It would also be interesting to attempt generalizations of our work to constructions involving supergroups as it was done for the
 $\l$-deformations in \cite{Hollowood:2014rla,Hollowood:2014qma}.

\no
It was shown in \cite{Vicedo:2015pna,Hoare:2015gda,Sfetsos:2015nya,Klimcik:2015gba,Klimcik:2016rov} that, the
$\l$- and $\eta$-deformations are related via Poisson-Lie T-duality
and appropriate analytic continuations. Our construction and results suggest that there likely exist new integrable $\s$-models
of the $\eta$-type to be constructed.
In that respect it would be interesting to study further the details of the algebraic and Hamiltonian structure of our theories.

\section*{Acknowledgements}

G. Georgiou would like to thank the
Physics Department of the National and Kapodistrian U. of Athens
for hospitality during this project. K.S. would like to thanks the Physics Division,
National Center for Theoretical Sciences of the National Tsing-Hua University in Taiwan
for hospitality and financial support during a stage of this work.

\end{document}